\documentclass[11pt]{article}
\usepackage{times}
\usepackage[height=7.6in,width=5.5in]{geometry}
\usepackage{graphicx}
\usepackage{amsmath,amsthm}
\usepackage{epstopdf}
\newtheorem{theorem}{Theorem}

\newcommand{\R}{{\bf R}}

\newcommand{\C}{{\bf C}}
\newcommand{\Z}{{\bf Z}}
\newcommand{\ra}{\rightarrow}

\newcommand{\ot}{[1,10)}
\newcommand{\ui}{[0,1)}

\begin{document}
\vspace*{5mm}
\begin{center} \Large \textbf{The Multiplication Game} 
\end{center}

\begin{flushright}
Kent E. Morrison \\
California Polytechnic State University \\
 San Luis Obispo, CA 93407\\
\tt{kmorriso@calpoly.edu}
  \end{flushright}
  
\large
\renewcommand{\baselinestretch}{1.17}   
\normalsize

{\renewcommand{\thefootnote}{}
\footnotetext{\emph{Math. Mag.} \textbf{83} (2010) 100--110.}}  

\section*{The game}
You walk into a casino, and just inside the main entrance you see a new game to play---the \emph{Multiplication Game}. You sit at a table opposite the dealer and place your bet. The dealer hits a button and from a slot in the table comes a slip of paper with a number on it that you cannot see. You use a keypad to choose a number of your own---any
positive integer you like, with as many digits as you like. Your number is printed on
the slip of paper along with the product of the two numbers. The dealer shows you the
slip so that you can verify that the product is correct. You win if the first digit of the
product is 4 through 9; you lose if it is 1, 2, or 3. The casino pays even odds: for a
winning bet of one dollar the casino returns your dollar and one more. Should you stay
and play?

It looks tempting. You you have six winning digits and the casino has only three!
But being skeptical, you take a few minutes to calculate. You write the multiplication
table of the digits from one to nine. Of the 81 products you see that 44 of them begin
with 1, 2, or 3, and only 37 begin with 4 through 9. Suddenly, even odds do not seem
so attractive! You abandon the game and walk further into the casino.

In the next room you find another table with the same game, but better odds. This
table pays \$1.25 for a winning one dollar bet. From your previous count you figure
that if the odds favor the casino by 44:37, then a fair payout would be 44/37 dollars
for a dollar bet; that is almost \$1.19, and this table is offering more. Should you stay
and play?

You open your laptop and write a computer program to count the products of the two
digit numbers from 10 to 99. (You realize immediately that in this game multiplying
by 1, 2, . . . , 9 is the same as multiplying by 10, 20, . . . , 90, and so you leave out the
one digit numbers.) You find that of these 8100 products the casino has 4616 winners
and you have 3484. The ratio $4616/3484$ is between 1.32 and 1.33, quite a bit more
than the \$1.25 being offered. You move on, heading to the back of the casino where
you might find the best odds.

Far back in a dark corner you find a high stakes table with the Multiplication Game.
This one offers to pay \$1.40 for a winning dollar bet with a minimum bet of \$100.
Now you take a little time to think it over. You run your computer program to multiply
all the three digit numbers between 100 and 999 and find that 461698 of the products
are winners for the casino and 348302 are winners for you. The ratio $461698/348302$
is 1.32557 to five decimal places. (The limit of this process, as the number of digits
increases, turns out to be about 1.32565.) The odds look good, so you stay to play.

You pick three digit numbers randomly. You win some and you lose some, but after
a hundred rounds you find yourself \$450 poorer. Why are you losing? Obviously the
casino is not choosing its numbers in the same way you are. If it were, you would be
ahead about \$320 by now. You wisely decide it is time to take a break from the table
and analyze the game more thoroughly.

\section*{Applying game theory}
 The Multiplication Game was first  described and analyzed by  B. Ravikumar \cite{Ravikumar07} as a two-person game in which the players choose $n$-digit integers for a fixed $n$. He determined the limit of the optimal strategy as $n$ goes to infinity. 
 
 In this article we have modified the game to allow positive integers of any length.
Note that it differs from most actual casino games (like blackjack or roulette) in that
both you and the casino can play strategically. The outcomes of all possible simultaneous
choices of the two players can be represented by an infinite matrix $A=(a_{ij})$
whose rows and columns are indexed by the positive integers. The rows of the matrix
correspond to your choices and the columns correspond to the choices of the casino.
Looking at the outcome from your point of view, we put 1 into the matrix at the $i j$
location if you win and 0 if the casino wins.
 
Clearly it is not in either playerÕs interest to choose the same number every time.
In the lexicon of game theory this would be called a \emph{pure strategy}. Instead the players
must use \emph{mixed strategies}, which are probability distributions on the set of positive
integers.

More formally, a mixed strategy is a probability vector $p=(p_1,p_2,\ldots)$, where each $p_i$ is non-negative and $\sum_i p_i =1$. It may also be helpful to think of p as a linear
combination of pure strategies, $p=\sum_i p_i \delta_i$, where $\delta_i$ is the pure strategy of choosing $i$ with probability 1. Thus, $\delta_i$ is the standard basis vector having 1 in the $i$th location and 0 everywhere else.

 Let $f(p,q)$ denote the probability that you win when the you use the mixed strategy
$p$ and the casino uses the mixed strategy $q$. Then
\[  f(p,q) =\sum_{i,j} a_{ij}p_i q_j .\]

(This is a doubly infinite sum, as each of $i$ and $j$ can be any positive integer.) You seek
to maximize your chance of winning, while the casino seeks to minimize it. That is,
you would like to find $p$ so that $f (p, q)$ is as high as possible while the casino chooses
$q$ to make it as low as possible.

(We are using $f (p, q)$ to represent a probability, rather than an expected profit,
because we want to be flexible about the payoffs. Your average profit per round at the
high stakes table is
\[ (+140) f (p, q) + (-100)(1-f (p, q)).\]
The adjustment required for other stakes is apparent.)

This leads to the standard definitions of the ``value'' of the game to each player. The value of the game to you is
\[ v_1= \sup_p \inf_q f(p,q), \]
and the value of the game to the casino is
\[ v_2 = \inf_q \sup_p f(p,q). \]

(To understand these expressions, first note that $\inf_q f (p, q)$ is the worst thing that can
happen to you if you choose strategy $p$. Therefore, $v_1$ is the best winning probability
that you can guarantee for yourself, independent of the casinoÕs choice. Similarly, $v_2$
is the best result---lowest winning probability for you---that the casino can guarantee
for itself.)

Now for any real valued function $f(u,v)$, whose domain is a Cartesian product $U \times V$, it is always the case that 
\[\sup_u \inf_v f(u,v) \leq \inf_v \sup_u f(u,v) .\] 
We leave the proof as an exercise for the reader. It follows that $v_1 \leq v_2$. 

For a finite game (one in which the sets of pure strategies for both players are finite)
we can apply the Minimax Theorem of von Neumann and Morgenstern. It states that \\
(a) \; the values are equal; that is, $ v1 = v2 = v$; and  \\
(b) \;  there are \emph{optimal strategies} $p$ and $q$ such that $v = \inf_q f (p, q) = \sup_p f (p, q)$.  \\
That means that if you choose $p$ and the casino chooses $q$, the result $v$ is guaranteed.
Unfortunately, the multiplication game is not a finite game. For infinite games the
Minimax Theorem is not true in general; there are examples of infinite games with
$v_1 \neq v_2$. In the analysis that follows we will show that the multiplication game is like
a finite game, at least to the extent that $v_1 = v_2$.

\section*{The analysis}
The analysis proceeds in steps.

First we observe that there is some redundancy in the set of pure strategies. For
example, the integers $21, 210, 2100,\ldots$ all give the same results when chosen by either
player. Thus, in the payoff matrix the rows representing these pure strategies are
identical, and so we can eliminate all but one of these rows. We can do the same with
the columns indexed by these equivalent choices.

Next we observe that the choices don't really need to be integers. A player could
choose 2.1 with the same result as choosing 21. We can define a reduced set of pure
strategies $X$ as the set of rationals in $[1, 10)$ having a terminating decimal expansion:
\[
  X= \Big\{ \sum_{k=0}^m d_k 10^{-k} | m \geq 0, d_k \in \{0,1,\ldots,9\}, d_0 \neq 0 \Big\}. 
  \]
Although the players are no longer choosing integers, we have not changed the game in any essential way, and we now have an efficient description of the infinite sets of pure strategies in which there is no redundancy. The players choose terminating decimals in the interval $\ot$ and the casino wins if the product begins with the digits 1, 2, or 3.

Next we modify the game again, this time possibly in an essential way. We enlarge
the sets of pure strategies to include all the real numbers in $[1, 10)$. Although this
increases the cardinality of the set of pure strategies from countable to uncountable,
the new game is easier to analyze and its solution will be essential in understanding
the original game. In a later section we will return to this modification to see whether
it affects our results.

So now the casino chooses $x$ and you choose $y$ in  $\ot$ with the casino winning if the first digit of $xy$ is 1,2, or 3. That is, the casino wins if
\[ 1 \leq xy < 4 \, \text{ or } \, 10 \leq xy < 40. \]
The points $(x,y)$ satisfying these inequalities make up the shaded region in Figure 1. The white region includes the points for which you win.
\begin{figure}

\centerline { \hspace{-0in}
\includegraphics[width=2.1in]{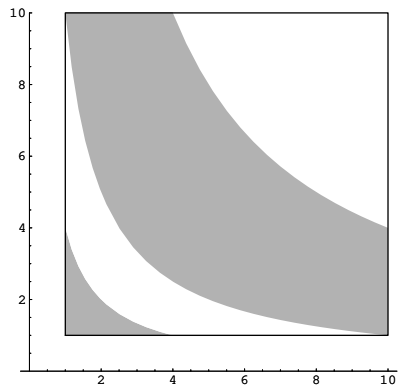}
}
\caption{Winning region for the casino}
\end{figure}

Finally, we straighten out the shaded region by using the logarithms of the numbers
rather than the numbers themselves. We let the casino choose $a = \log_{10} x$ and
you choose $b = \log_{10} y$. Clearly, choosing $a$ and $b$ is equivalent to choosing $x$ and $y$.
(However, as we will see, mixed strategies can look very different when described in
terms of $a$ and $b$.) Now the casino wins when
\[ 0 \leq a+b \le \log_{10} 4 \, \text{ or } \, 1 \leq a+b \le 1 + \log_{10}4, 
\]
which is equivalent to
\[  a + b \bmod 1 \in [0, \log_{10} 4) .\]
Figure 2 is a complete picture of the game---essentially the payoff matrix. You choose a horizontal line and independently the casino chooses a vertical line. You win if the point of intersection is in the white region. 
\begin{figure}[t]
\centerline { \hspace{-0in}
\includegraphics[width=2.1in]{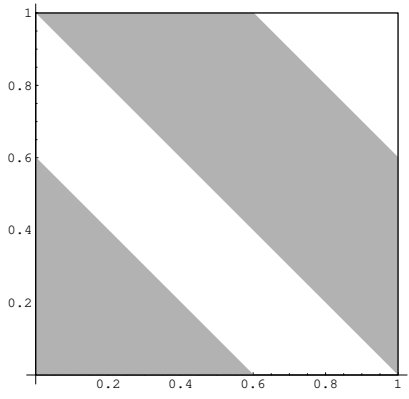}
}
\caption{Winning region for the casino}
\end{figure}

What are mixed strategies now that the players have an uncountable number of choices? They should be probability distributions on the set $\ui$ of pure strategies.

There are many ways to describe probability distributions on an interval. Density
functions and cumulative distribution functions (cdf's) come to mind. Most formally, a
probability distribution is any measure on the interval, which is a non-negative function
defined on a suitable collection of subsets of $[0, 1)$. The measure must be countably
additive and assign the value 1 to the whole interval. The mixed strategies we need can
be described in terms of density functions, while pure strategies are discrete distributions
concentrated at single points.

\section*{Solving the (modified) game}
In order to motivate the solution of this game, consider for a moment an analogous finite game. Looking at Figure 2 one can see that the same proportion of each horizontal line lies within the shaded region, and the same is true for the vertical lines. A finite game with a similar structure is one in which both players have the same set of pure strategies $\{1,\ldots,n\}$; all entries in the payoff matrix are 1 or 0, and there are the same number of 1's in each row and in each column. The payoff matrix, then, is something like a discrete version of Figure 2 with the location of the 1's playing the role of the white region.  

For a game with these properties we claim that an optimal strategy for both players is the uniform probability distribution 
\[ (1/n,\ldots, 1/n). \]
To see this, let $p$ be uniform. No matter which $j$ the column player chooses, the row player wins with probability
\[  \sum_i a_{ij} p_i = \sum_i a_{ij} (1/n)=\frac{1}{n} \sum_i a_{ij} = \frac{c}{n}, \]
where $c$ is the number of 1's in each row. Thus, $v_1 \geq c/n$.
On the other hand, if the column player uses the uniform strategy for $q$ and the row player chooses any row $i$, then the row player wins with the same probability
\[ \sum_j a_{ij} q_j=\sum_j a_{ij} (1/n)= \frac{1}{n} \sum_j a_{ij} = \frac{c}{n},\]
and thus $v_2 \leq c/n$. Since $v_1 \leq v_2$, this proves that $v_1=v_2$ and from that it follows that the uniform probability distributions are optimal. 

With this analogy to guide us, consider what happens when you choose your number uniformly in $\ui$. This means your mixed strategy is the uniform distribution on this
interval, which we denote by $\lambda$. Assume that the casino uses the pure strategy $\delta_a$. Thus, the outcome is on the vertical line $\{(a,b) | b \in \ui \}$ and the probability that the point $(a,b)$ is in the shaded region is $\log_{10}4$. That is the casino's probability of winning. Your probability of winning is $1-\log_{10}4$, regardless of the value of $a$, and so $v_1 \geq 1- \log_{10}4.$

Now we look at the game from the casinoÕs point of view and reason in the same
way. No matter what pure strategy $\delta_b$ that you employ, the casino can choose its number
uniformly and win with probability $\log_{10} 4$. Thus, the casino can guarantee winning
with at least this probability no matter what you do, and so 
$v_2 \leq  1- \log_{10} 4$. Since
$v_1 \leq v_2$, we see that they are in fact equal, and it follows that the uniform distribution is
optimal for both players. These are the conclusions that the MiniMax Theorem would
have given us if it had been applicable. Since $\log_{10}4 \approx 0.60206$, the casino will win just over 60\% of the time, which makes the odds a bit higher than $3:2$. To make the game fair the casino should pay you a bit more than \$1.50 for your winning one-dollar bet. The exact amount is 
\[  \frac{\log_{10}4}{1-\log_{10}4}  \approx 1.5129 ,\]
 but the casino was only paying \$1.40 in the high stakes game, and so you found yourself losing after playing for a while.

Now we transfer the uniform measure $\lambda$ on the logs in $\ui$  back to a measure on $\ot$ that we denote by $\beta$. Thus $\beta$ assigns to the interval between $x_1$ and $x_2$ the probability   $\log_{10}x_2 -\log_{10}x_1$. This probability measure $\beta$ has the density function
\[  f(x)=\frac{1}{\ln 10} \frac{1}{x} \]
shown in Figure 3.
The area between $x_1$ and $x_2$ and lying under the graph of $f$ gives the probabilty that $x$ is between $x_1$ and $x_2$.
\begin{figure}
\centerline {\hspace{-0in}
\includegraphics[width=3in]{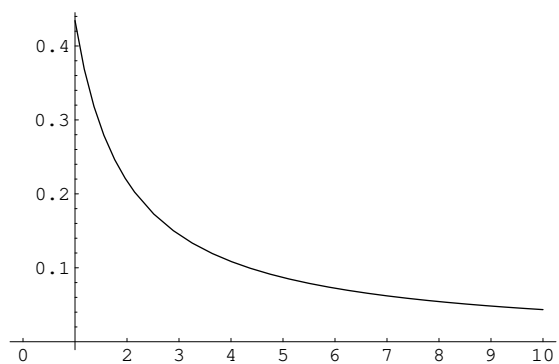}
}
\caption{Plot of the Benford density $f(x)=\frac{1}{(\ln 10) x}$}
\end{figure}

In recent years this logarithmic probability distribution has become known as the \emph{Benford distribution}, named for the physicist Frank Benford who investigated the relative frequency of leading digits of more than 20,000 numbers in several datasets from diverse sources such as populations of cities and the areas of river basins. 
Although Benford described the phenomenon over 70 years ago \cite{Benford38}, the first discovery was actually due to the astronomer and mathematician Simon Newcomb who observed the phenomenon more than half a century earlier \cite{Newcomb81}. Newcomb's paper did not spark any further work in the years that followed. Benford, however, was fortunate to have his paper appear just in front of an influential paper in modern physics having Hans Bethe as one of the authors, and so it was widely seen by other scientists. This helped to attach Benford's name to the empirical observations and began what has become a small industry; there are now more than 700 papers in a comprehensive Benford online bibliography \cite{Berger&Hill09}. (Naming the phenomenon after Frank Benford illustrates \emph{Stigler's Law of Eponymy}, which states that no scientific discovery is named for its discoverer. Appropriately, Stigler attributes his eponymous law to Robert Merton.) Benford's Law describes not just the distribution of the first significant digit but also the distribution of all significant digits. In its general form the law is the logarithmic distribution on the set of real numbers between 1 and 10, and even more generally the number base can be any positive integer $b \geq 2$, in which case the Benford distribution is supported on the interval $[1,b)$. The special cases concerning any particular significant digits can be derived from the continuous distribution. For enlightening accounts of Benford's Law we recommend the articles by Raimi \cite{Raimi69}, Hill \cite{Hill95b,Hill98}, and Fewster \cite{Fewster09} or the statistics text by Larsen and Marx \cite{Larsen-Marx05}.

What happens if we change the game so that the casino wins if the first digit is prime? Or change the game so that the winner is determined by the value of the second digit of the product? To what extent can we solve the game, i.e., determine the optimal strategies and the value of the game, when the winning conditions are changed?

Let $W \subset \ot$ be the \emph{winning set} for the casino; the casino wins when the product $xy$ lands in $W$. For the version we have analyzed the winning set $W$ is the interval $[1,4)$, consisting of the numbers whose first digit is 1, 2, or 3. Now look at the logs of the numbers in $W$ and call that set $Z$. If the logs are $a$ and $b$, then the casino wins when $a + b \pmod 1$ is in $Z$. Now plot the region in the unit square consisting of the points $(a,b)$ for which the casino wins. See Figure 4 for the case in which the casino wins when the first digit is prime. In general this region consists of bands between lines having slope $-1$. Each horizontal line and each vertical line meets the shaded region in the same proportion. The points on the horizontal axis that are in the shaded region are $(a,0)$ where $a \in Z$. Thus, the proportion of each line that lies in the shaded region is $\lambda(Z)$, the Lebesgue measure of $Z$, which is the same as $\beta(W)$, the Benford measure of $W$. Using the same reasoning as before we conclude that if the casino chooses its logarithm uniformly, then it will win with probability $\lambda(Z)$ regardless of what you do, and if you choose your logarithm uniformly you will win with probability equal to $1-\lambda(Z)$.

\begin{figure}
\centerline { \hspace{-0in}
\includegraphics[width=2.1in]{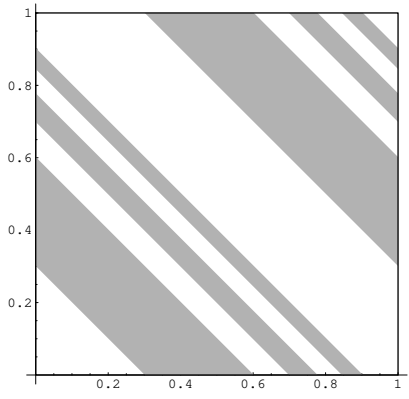}
}
\caption{Winning region for the casino for prime first digit}
\end{figure}

Just how far can we push this approach? Certainly $W$ (or $Z$) can be any finite union of intervals but we can even allow countable unions of intervals; it does not matter whether they are open, half-open, or closed. But non-measurable sets are too strange to be used for winning sets, because if $Z$ is not Lebesgue measurable, then we cannot make sense of the statement that every horizontal and vertical line meets the set $\{(a,b) | a + b \pmod{1} \in Z\}$ in the same proportion.
We summarize this discussion in the following theorem.

\begin{theorem}
Let the casino's winning set $W \subset \ot$ be a finite or countable union of intervals. Then an optimal mixed strategy for both you and the casino is to choose your numbers from the Benford distribution on $\ot$, or, equivalently, to choose the logarithms uniformly in $\ui$. The probability that the casino wins is $\beta(W)$.
\end{theorem}

\section*{Solving the original game}
Now we return to the analysis of the original game in its reduced form in which you and the casino choose numbers in the set $X$ of terminating decimals in $\ot$. We will show that by approximating the Benford distribution we can find mixed strategies that come arbitrarily close to being optimal. That is, we will show that 
\[\sup_p \inf_q f(p,q)=\inf_q \sup_p f(p,q)=\log_{10}4,\]
but we will not actually find strategies that attain this. 

Let $X_n$ be the subset of $X$ whose elements have a terminating expansion with $n$ digits,
\[
  X_n= \bigg\{ \sum_{k=0}^{n-1} d_k 10^{-k} \, | \, d_k \in \{0,1,\ldots,9\}, d_0 \neq 0 \bigg\}.
  \]
Consider the following strategy. Fix a positive integer $n$. Generate a random number  $a \in \ui$, compute $10^a$ and then choose $x \in X$ to be the nearest $n$-digit number less than or equal to $10^a$. This defines a probability measure $\beta_n$ that is concentrated on the finite set $X_n$.
The probability mass at the point $x \in X_n$ is given by 
\[
\beta_n\{x\}= \log_{10}(x+1/{10^{n-1}}) - \log_{10} x .
\]
Let $F(x)=\log_{10}x$ be the cumulative distribution function (cdf) of $\beta$ and $F_n$ the cdf of $\beta_n$. Then $F_n$ has jumps at the points in $X_n$ and is always greater than or equal to $F$. (See Figure 5.) The maximum difference between $F_n$ and $F$ occurs at $x=1$, where 
\begin{equation} \label{cdfdiff}
 F_n(1) - F(1)= \log_{10}( 1 + 1/10^{n-1}) -0 < \frac{1}{10^{n-1}}.
\end{equation} 
\begin{figure}
\centerline {
\includegraphics[width=3in]{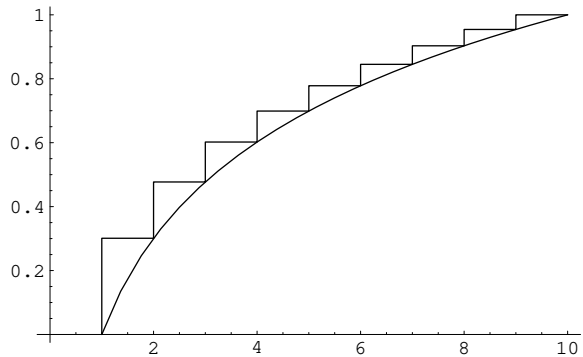}
}
\caption{The cdf's of $\beta_1$ and $\beta$}
\end{figure}

Now if the casino uses $\beta_n$ for its mixed strategy and you play $y \in X$ (or, for that matter, any $y \in \ot$), then the casino will win with probability 
$\beta_n(V_y) $ where 
\[ V_y=\{x \in X \,| \, xy \in [1,4) \cup [10,40) \} \label{vy} \]
 If $\epsilon > 0$, then for $n$ sufficiently large we have $|\beta_n(V_y)-\beta(V_y) | < \epsilon$, and this estimate holds for all $y$ because $V_y$  is either an interval or a union of two intervals in $\ot$ and the measures of $V_y$ can be expressed with the cdf's $F_n$ and $F$ evaluated at the endpoints of the intervals where they differ by at most $10^{n-1}$ according to  (\ref{cdfdiff}). Therefore, the casino can guarantee a win with probability at least $\log_{10}4 - \epsilon$, and so $v_1 \geq 
\log_{10}4 - \epsilon$.

Similar reasoning shows you can use $\beta_n$ for $n$ sufficiently large and guarantee that the casino wins with probability no more than $\log_{10} 4 + \epsilon$. Therefore, $v_2 \leq \log_{10}4 - \epsilon$. These inequalities hold for all $\epsilon$, and thus $v_1=v_2=
\log_{10}4$. It is worth repeating that we have not found an optimal strategy that actually achieves the value but rather a family of strategies that come arbitrarily close to optimal. It is tempting to consider the limit of the $\beta_n$, which is $\beta$, as an optimal strataegy, but $\beta$ is not a probability distribution on $X$.
If an optimal strategy $\mu$ exists as a probability measure on $X$, then it must have the property that $\mu(V_y)=\log_{10}4$ for all $y \in X$. There is no evident way to produce such a measure even if one exists. (Note added after publication: Kenneth Ross has communicated to the author a proof that no such measure exists.)

\section*{The group game}
How far we can generalize the multiplication game? Although there may be other paths to follow, we will assume that there is a binary operation that combines the players' choices to produce the result.  Does the operation need to be associative? Commutative? Have an identity? Inverses? And what about the nature of the set on which the operation is defined?

\newcommand{\mg}{\R_+/\langle 10 \rangle}
In the original game we use the positive integers with multiplication. The operation is associative, commutative, and has an identity, but does not have inverses. The same is true of the reduced version with the set $X$, where after multiplying we move the decimal point if necessary to get a number between 1 and 10. However, by extending the set of pure strategies to be all real numbers between 1 and 10, we get inverses, and the result is that we have a group. This group is nicely described as the quotient group $\R_+/\langle 10 \rangle$, where $\R_+$ is the multiplicative group of positive real numbers and  $\langle 10 \rangle$ is the subgroup generated by 10 consisting of all integral powers of 10. The numbers in $\ot$  are unique coset representatives of the subgroup $\langle 10 \rangle$. With logarithms we use the set $\ui$ with addition mod 1, which is another way to describe the  quotient group $\R/\Z$, where $\R$ is the additive group of real numbers and $\Z$ is the integer subgroup. An isomorphism from $\R/\Z$ to $\mg$ is given by $a \mapsto 10^a$.
For the $\R/\Z$ game we proved that Lebesgue measure is optimal,  and for the $\R_+/\langle 10 \rangle$ game it is the Benford measure that is optimal. These measures are special for their respective groups in that they are \emph{invariant}. For Lebesgue measure it means that $\lambda(E)=\lambda(a+E)$ for a subset $E$ of $\ui$ and for $a \in \ui$. For the Benford measure it means that $\beta(E)=\beta(xE)$ for $E \subset \ot$ and $x \in \ot$. (Addition and multiplication must be done in the quotient groups.) Furthermore, $\lambda$ and $\beta$ are probability measures, meaning that they are positive measure with total mass equal to one.

There is a class of groups having exactly the properties necessary to generalize Theorem 1, namely  the class of \emph{compact topological groups}. Among these groups are the groups we have just described, $\R/\Z$ and $\R_+/\langle 10 \rangle$, both of which are abelian and topologically equivalent to circles. Also, every finite group (abelian or not) is a compact topological group with its discrete topology. For infinite non-abelian examples there are the groups of isometries of $\R^n$ for $n \geq 2$. By contrast, topological groups that are not compact include the real numbers under addition, the non-zero real numbers under multiplication, the invertible $n \times n$ matrices over $\R$ or over $\C$, and any infinite group  with the discrete topology (such as the integers under addition). 

In general, a topological group is a topological space $G$ together with a continuous group operation $G \times G \ra G: (g_1,g_2) \mapsto g_1g_2$ and a continuous inverse map $G \ra G: g \mapsto g^{-1}$. With compactness comes the existence of a unique invariant (under left and right multiplication) probability measure $\lambda$ known as  \emph{Haar measure} \cite{Halmos50}.  For finite groups Haar measure is simply normalized counting measure, whereas for $\R/\Z$ it is Lebesgue measure and for $\mg$ it is the Benford measure. 

With a compact topological group $G$ we can generalize Theorem 1 as follows. Let $W \subset G$ be a $\lambda$-measurable subset. The casino chooses $x \in G$, you choose $y \in G$, and the casino wins if $xy$ is in $W$. Then for both players an optimal mixed strategy is to use Haar measure $\lambda$, and the value of the game, i.e., the probability that the casino wins, is $\lambda(W)$. The proof, the details of which will be omitted, uses the invariance properties of $\lambda$ to show that when the casino uses $\lambda$ it does not matter what strategy you use, and when you use $\lambda$ it does not matter what the casino does. 

If any of the hypotheses are relaxed, then we do not have complete solutions to the game. The analysis becomes more difficult, something we have already seen with the original game played on the positive integers or in the equivalent version on the terminating decimals. Two important properties are lacking: inverses and compactness. We can exhibit mixed strategies arbitrarily close to optimal, and thus show that the game has a value, but we cannot exhibit strategies that actually achieve the optimum. Even if we add inverses to the set of terminating decimals by including all rational numbers in $\ot$, the resulting group is not compact. It is countable and infinite and so it cannot carry an invariant probability measure because each element of the group would have the same non-zero mass and so the total mass would be infinite. We doubt that an optimal strategy exists but it is a question we leave unanswered. 

\vspace{3mm}
\noindent \textbf{Acknowledgment.} 
\; The author would like to thank B. Ravikumar, inventor of the multiplication game, for providing a copy of his paper describing and analyzing the game. The limiting distribution that he found is, of course, the Benford distribution.

\vspace{3mm}
\noindent \textbf{Summary.} 
\; The Multiplication Game is a two-person game in which each player chooses a positive integer
without knowledge of the other playerÕs number. The two numbers are then multiplied together and the first digit of the product determines the winner. Rather than analyzing this game directly, we consider a closely related game in which the players choose positive real numbers between 1 and 10, multiply them together, and move the decimal point, if necessary, so that the result is between 1 and 10. The mixed strategies are probability distributions on this interval, and it is shown that for both players it is optimal to choose their numbers from the Benford distribution. Furthermore, this strategy is optimal for any winning set, and the probability of winning is the Benford measure of the playerÕs winning set. Using these results we prove that the original game in which
the players choose integers has a well-defined value and that strategies exist that are arbitrarily close to optimal. Finally, we consider generalizations of the game in which players choose elements from a compact topological group and show that choosing them according to Haar measure is an optimal strategy.
\pagebreak

\end{document}